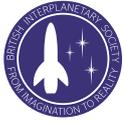

# The Spider Stellar Engine: a Fully Steerable Extraterrestrial Design?

**CLÉMENT VIDAL**  Center Leo Apostel, Vrije Universiteit Brussel, Belgium

**Email** contact@clemvidal.com

**DOI** https://doi.org/10.59332/jbis-077-05-0156

A long-lived civilization will inevitably have to migrate towards a nearby star as its home star runs out of nuclear fuel. One way to achieve such a migration is by transforming its star into a stellar engine, and to control its motion in the galaxy. We first provide a brief overview of stellar engines and conclude that looking for technosignatures of stellar engines has taken two roads: on the observational side, hypervelocity stars have been the target of such searches, but without good candidates. On the theoretical side, stellar engine concepts have been proposed but are poorly linked to observable technosignatures. Since about half the stars in our galaxy are in binary systems where life might develop too, we introduce a model of a binary stellar engine. We propose mechanisms for acceleration, deceleration, steering in the orbital plane and outside of the orbital plane. We apply the model to candidate systems, spider pulsars, which are binary stars composed of one millisecond pulsar and a very low-mass companion star that is heavily irradiated by the pulsar wind. We discuss potential signatures of acceleration, deceleration, steering, as well as maneuvers such as gravitational assists or captures.

## 1    INTRODUCTION

Why would a long-lived civilization even engage in interstellar travel? After all, interstellar distances are immense, the travel times typically exceed human lifetimes, and require significant economical and political investment to secure the required technology and energy.

Yet, two universal evolutionary motivations will make interstellar travel a necessity to any long-lived civilization: *survival* and *reproduction*. The survival motivations include avoiding a death threatening supernova, migrating towards a nearby star as the home star fades away (Zuckerman 1985 [1]; Hansen and Zuckerman 2021 [2]; Hansen 2022 [3]), or even securing the long-term access to a maximum of stars before they become permanently inaccessible due to the accelerated expansion of the universe (Hooper 2018 [4]). To avoid frequent migrations, an advanced civilization might also choose to target high stellar density environments such as globular clusters or the galactic center.

The reproduction motivations would include engaging in interstellar life spreading, also known as directed panspermia (Crick and Orgel 1973 [5]; Mautner 1997 [6]). Secondary motivations may include the will and drive to expand; pursuing more ambitious galactic engineering projects  (e.g. Badescu, Cathcart, and Schuiling 2006 [7]; Vidal 2019 [8]) or simply curiosity to explore the galaxy.

However, the tyranny of the rocket equation and the large interstellar distances still require careful considerations to answer the payload question: what should a civilization bring? The two extreme options are either to send a *very light* payload that could be accelerated to relativistic velocities, which is the philosophy behind Breakthrough Starshot project. The other extreme option is to travel very *heavy* by moving the stellar system as a whole, with the help of a *stellar engine*.

At first, the option of a stellar engine sounds like science fiction, but it presents many advantages: a civilization traveling with its star maintains its source of energy, and thus can continue its normal activities: there is thus no need to rush, and a stellar engine may be the fastest *evolutionary* way to travel, in the sense that evolution and complexity do not have to re-start almost from scratch as is the case with directed panspermia, or with space colony projects.

Such a vision of controlling larger and larger scale processes stems from a long tradition in SETI, from Kardashev's (1964) [75] spatial and energetic scale (see Ćirković 2015 [9] for an overview) and related variants extrapolating the ability to move larger and larger objects (Romanovskaya 2022 [10]), or extrapolating waste management at larger and larger scales (Vidal and Smart 2024 [11]).

   So far the literature on stellar engine designs has only considered how to move a single star, and not a binary star system. However, we know that about half of star systems are binaries

---

Paper presented at the 8th Interstellar Symposium of The Interstellar Research Group.

(Han et al. 2020 [12]), and that they are capable of hosting habitable worlds (Eggl 2018 [13]). On the technosignature side, various hypotheses have been proposed in relation to binary stars (Fabian 1977 [14]; Corbet 1997 [15]; Vidal 2016 [16]; 2019 [8]; Lacki 2020 [17]). In this context, it makes sense to pay attention to the problem of designing a binary stellar engine and to note that it is different from a single star stellar engine, mostly because it adds the constraints of the orbital nature of the system, and that more than ~1 $M_\odot$ is involved.

We first present a quick overview of previous stellar engine designs (Section 2), and then introduce the first binary stellar engine model, including mechanisms to accelerate, decelerate, steer in the orbital plane and steer outside the orbital plane (Section 3). We then show that a small subset of binary stars, spider pulsars, may be interpreted as *spider stellar engines* in action. We substantiate this interpretation by giving candidate examples, and propose to contrast testable astrophysical and extraterrestrial intelligence (ETI) explanations and predictions.

In sum, this paper presents a first binary stellar engine design, thereby widening the class of stellar engine possibilities, and proposes testable and observable stellar engine technosignatures to further probe the spider stellar engine hypothesis.

## 2    STELLAR ENGINES

The great science fiction writer Olaf Stapledon (1937/1953) pioneered the idea of the stellar engine by envisioning to form a small artificial Sun to sustain long interstellar voyages back in 1937 [18]. Twenty years later, the great Swiss astronomer Fritz Zwicky (1957/2012, 260) considered seriously the possibility to use Sun and planets as nuclear propellants [19]:

> "Considering the Sun itself, many changes are imaginable. Most fascinating is perhaps the possibility of accelerating it to higher speeds, for instance 1,000 km/s directed toward α-Centauri in whose neighborhood our descendants then might arrive a thousand years hence. All of these projects could be realized through the action of nuclear fusion jets, using the matter constituting the Sun and the planets as nuclear propellants."

The first detailed stellar engine model was introduced by Shkadov in 1988 [20], proposing to place a non-orbiting giant parabolic mirror held at fixed distance above the Sun in order to create thrust. However, the Shkadov thruster design has been qualified as a passive stellar engine [21], because it creates thrust by merely reflecting sunlight. Building on the landmark study of the long-term future of civilization in our solar system by Criswell in 1985 [22], Martyn Fogg (1989) did propose the first active stellar engine that involves actively ejecting mass from the Sun in order to create thrust [23].

More recently, Caplan (2019 [24]) did a detailed study and update of the Shkadov thruster (see also Svoronos and Caplan 2021 [25]), and did propose a new active thruster design using a thermonuclear ramjet. However, one issue is that the engine has to generate two opposite directions of thrust: one direction away from the star to generate thrust, but also another thrust towards the star to remain stable. Svoronos (2020, 306 [26]) noticed that it is energetically expensive, and concluded that "less than half of the thrust generated by the engine is used to accelerate the star."

Svoronos' (2020 [26]) "star tug" design aims to remedy the issue by keeping the stellar engine gravitationally bound to the star, inspired in parts by gravity tractors techniques to deflect asteroids (Lu and Love 2005 [27]; Mazanek et al. 2015 [28]). To solve the issue of counteracting the gravitational pull of the star, Svoronos also noted the possibility that "the engine, or many copies of the engine, could be placed in a stable orbit around the sun, thereby counteracting the force of gravity, and the engines could perform pulsed bursts of thrust once each orbit when they are at the appropriate position." As we will see, this is a key insight for a binary stellar engine.

On the observational side, since stars moving at relativistic speeds seems very unlikely to happen naturally, Lingam and Loeb (2020 [29]) looked for hypervelocity stars in the Gaia catalog that could be interpreted as stellar engine technosignatures. They did not find good candidates. Forgan (2013 [30]) proposed methods for detecting Shkadov thrusters during transits, although he concludes that it remains challenging. In sum, as of today there are no candidate examples of stellar engines in observational data.

## 3    A BINARY STELLAR ENGINE MODEL

Here we propose a model of a binary stellar engine, inspired both by existing stellar engine designs reviewed above, and by the phenomenology of a subclass of binary millisecond pulsars (MSPs), spider pulsars (see also section 4). We assume that the payload is a compact object (neutron star) of about 1.8 $M_\odot$, and the propellant is its low-mass companion star (0.01-0.7 $M_\odot$).

We now explain the four steering mechanisms for acceleration, deceleration, steering the direction in the orbital plane, and steering out of the orbital plane that are illustrated in Fig 1.

### 3.1    Acceleration and deceleration

The first key challenge for any engine is to effectively generate thrust. By definition, thrust is a function of both the *speed* at which propellant is expelled, and the *amount of propellant*. Thrust requires not only expelling material, but expelling it out of the gravitational bound of the binary system. One can make a rough Newtonian estimate of the escape velocity necessary for material to be expelled out. Assume an orbital separation between the neutron star and the companion $a = 10^9$ m, and assume that the system as a whole can be assimilated to one single object of radius $R = a = 10^9$ m and a mass of $M = 2.14\ M_\odot$, we can use the formula:

$$v_{e\_min} = \sqrt{\frac{2GM}{R}} \qquad (1)$$

that gives a minimum escape velocity $v_{e\_min}$ of 753 km.s$^{-1}$. Following Newton's third law, ejected material must reach this minimal velocity to have any effect on the motion of the system. Note that this estimate neglects the binarity and other dynamics of the system that may give rise to slightly different results.

If we want to further compute the acceleration capacity and the Δv associated with the rocket equation:

$$\Delta v = v_e \ln \frac{m_0}{m_f} \qquad (2)$$

we need to assume an evaporation rate, and a duration of evaporation. We propose to use the assumptions from Fogg [23], i.e. an evaporation rate of $3 \times 10^{-9}\ M_\odot$.y$^{-1}$ during 10.7 million years. Using $m_o = 2.14\ M_\odot$ and $m_f = (3 \times 10^{-9} \times 10.7)$, this results in Δv = 11.4 km.s$^{-1}$ and an acceleration capacity of $3.4 \times 10^{-11}$ m.s$^{-1}$.

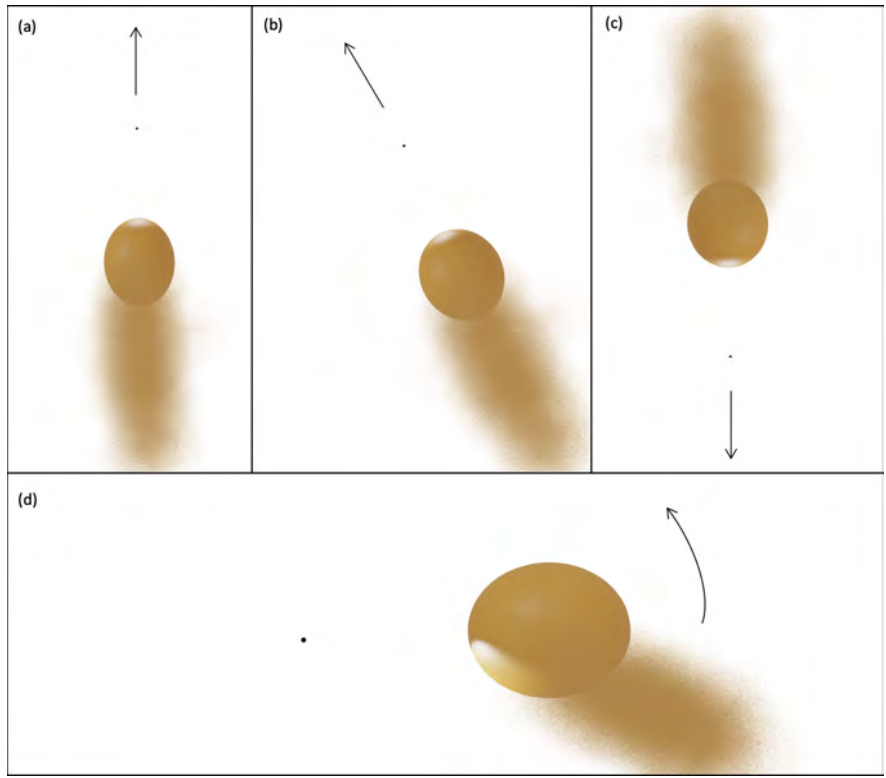

**Fig.1** Four steering configurations of the binary stellar engine. Figures (a-c) are top views, face on, while figure (d) is side on. In situation (a) the stellar engine is accelerating or cruising; in situation (b), assuming the system velocity is towards the top, the thrust creates a force towards the left; in situation (c), assuming the system velocity is towards the top, the thrust creates a decelerating force. Situation (d) changes the orbital plane by asymmetric heating of the companion, which creates a lifting force in relation to the orbital plane. Note that the pulsar size and orbital separation are not to scale.

However, these values need to be significantly lowered down because thrust happens at most once per orbit, when the companion is at an appropriate orbital phase to be evaporated. However, we assume an extremely compact binary, with orbital periods on the order of a few hours, so the thrust would still be very consistent over time. The duration of the evaporation itself defines the equivalent of the shape of the cone of a rocket nozzle. A long evaporation defines a widely open nozzle angle, while a short evaporation – possibly as short as the companion's radius eclipsing the pulsar – would define the narrowest nozzle angle. Note that the resulting pulsed thrusting also exists in spacecraft propulsion designs, with electromagnetic pulsed inductive thrusters, electrothermal pulsed plasma thrusters, or nuclear pulse propulsion.

If we assume that the duty cycle of the engine is 10% of the orbit, then the evaporation would be 10% of $3 \times 10^{-9}$ $M_\odot.y^{-1}$ which lowers $\Delta v = 1.13$ km.s$^{-1}$ while the acceleration capacity becomes $3.3 \times 10^{-12}$ m.s$^{-1}$. In comparison to other stellar engine designs (see Table 1), these figures mean that the engine is much less capable given the high payload and reduced duty cycle.

Up to now, we did all our calculations with the minimal escape velocity $v_{e\_min}$, but considering that pulsar winds typically display synchrotron emission (e.g. Harding and Gaisser 1990 [31]; Arons and Tavani 1993 [32]), wind particles reach commonly relativistic velocities. If we assume that the ejection of matter can reach such relativistic velocities, e.g. $v_{e\_max} = 0.75\,c$ then $\Delta v = 337$ km.s$^{-1}$ and the acceleration is $10^{-9}$ m.s$^{-1}$. This gives an upper limit for the binary stellar engine.

**TABLE 1: Comparison of key features and capacities of stellar engines**

| Feature -> Stellar engine model | Class | Distance from the star | Acceleration capacity (m.s$^{-2}$) | Δv (km.s$^{-1}$) | Remarks |
|---|---|---|---|---|---|
| Shkadov thruster | Passive | $1.5 \times 10^{11}$ | $10^{-12}$ | 200 (max) | This is the first stellar engine model; the Δv was estimated by Caplan (2019 [24]). |
| Fogg thruster | Active | Unspecified | $6 \times 10^{-11}$ | 20 | The acceleration is computed in the present paper, with the assumptions in Fogg's (1989 [23]) paper of $3 \times 10^{-9}$ $M_\odot.y^{-1}$ of propellant, and 10.7 million years. |
| Caplan thruster | Active | $6 \times 10^{10}$ m | $10^{-9}$ | 340 | Assuming a quarter of the star is used as propellant. |
| Star Tug | Active | Unspecified | min: $7.2 \times 10^{-9}$ max: $2.3 \times 10^{-6}$ | n/a | The acceleration capacity varies depending on assumptions described in Svoronos 2020 [26]. |
| Binary Stellar Engine | Active | $10^9$ m | min: $3.3 \times 10^{-12}$ max: $10^{-9}$ | min: 1.13 max: 337 | We assume an evaporation rate of $3 \times 10^{-10}$ $M_\odot.y^{-1}$ during 10.7 million years, a total initial mass of 2.14 $M_\odot.y^{-1}$ to move, and an exhaust velocity of 753 km.s$^{-1}$. The upper bound (max) uses the same numbers except with a relativistic exhaust velocity of 0.75 $c$. |

The question of how precisely to evaporate the companion is not trivial to model, but we can make the following remarks. First, the engine is best positioned very near the companion star to have more thrust (Svoronos 2020 [26]), so here again a close binary would work better than a wide binary. Also, the companion star needs to be heavily heated, especially in its upper layers, possibly with X-ray or gamma-ray radiation before the material starts to be evaporated, and can be channelled with magnetic fields to reach escape velocities. The detailed modeling of such processes is highly technical and outside the scope of this paper, although we will discuss some of these issues in the next Section 4.

### 3.2  Deceleration

Since space is mostly empty, without friction, it is almost impossible to brake. The simplest way to decelerate is thus to produce an active thrust in the opposite direction of motion (see Fig. 1c). Additional drag can be obtained with a passive method, by deploying a magnetic sail from the neutron star and starting to transfer the system's momentum to the interstellar medium (see e.g. Gros 2017 [33] on magnetic sails). We illustrate the deployment of such a magnetic sail around the companion star in Fig. 2a. To switch back to an acceleration or cruising phase, the magnetic sail could be retracted in order not to generate drag. However, in that case the magnetic field of the companion star would become stronger than the magnetic sail, and the shock would wrap around the neutron star (see Fig. 2b). The detailed modeling of a magnetic sail in the context of the binary stellar engine remains to be further elaborated.

### 3.3  Steering

Let us now tackle the issue of steering. To steer a spacecraft, one typically needs to be able to control *yaw* (moving the nose from side to side), *pitch* (moving the nose up or down), and *roll* (circular movement as the spacecraft moves forward).

The issue of controlling yaw is solved by precisely timing the evaporation at various orbital phases (Fig 1b). To choose a direction, it suffice to evaporate the companion star once per orbit, at a specific orbital phase in order to create consistent thrust in one direction. Of course, if the irradiation would be continuous along the orbit, the thrust would be omnidirectional and thus no overall directional momentum would be gained, a bit like when a car is drifting in a full circle.

By evaporating at different orbital phases, one can obtain directional thrust towards any desired direction on the orbital plane. As we mentioned earlier, the idea of positioning in a stable orbit and triggering pulsed bursts of thrusts once per orbit was suggested by Svoronos (2020 [26]) as a way to improve on Caplan's (2019 [24]) model, a steering solution that we may call 360° thrust vector control.

In the context of a binary stellar engine, the issues of *roll* and *pitch* are the same problem: changing the orbital plane. A simple solution is to irradiate and evaporate the companion star in an asymmetric way, to generate a force that pushes the companion star towards a new desired orbital plane as illustrated in Fig. 1d.

### 3.4  Discussion

We can add a few more issues regarding the binary stellar en-

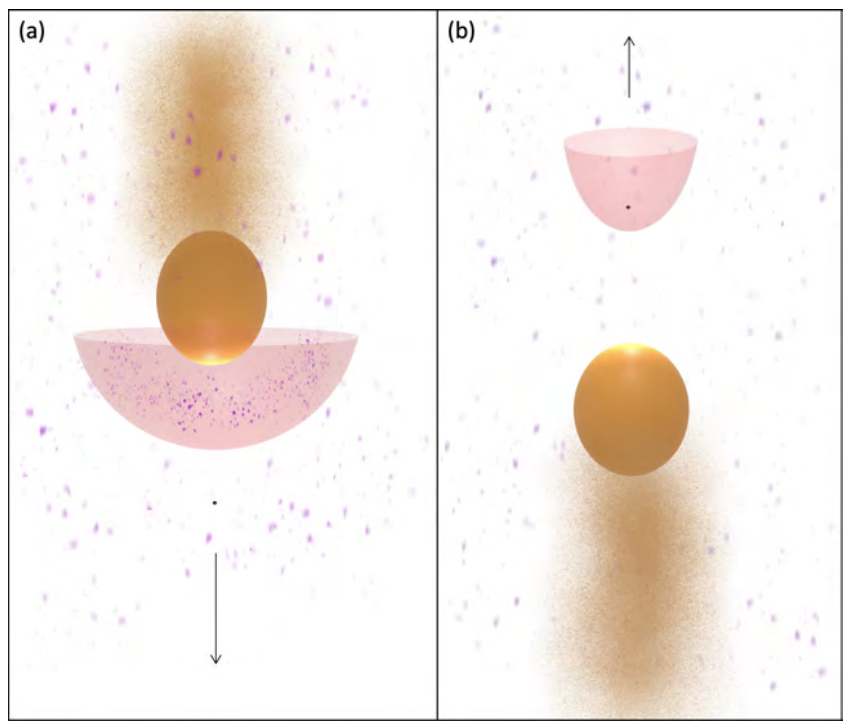

**Fig.2** (a) Deceleration is improved by using both an active thrust, and a passive collection of interstellar medium particles (in magenta) through a magnetic sail (in pink). (b) In an acceleration or cruising phase, the magnetic sail is retracted and a smaller collecting area results.

gine. The first is that evaporation creates a force on the companion star, and therefore modifies the orbital separation. In the case of acceleration (Fig. 1a), the orbital separation will tend to extend, while in the case of deceleration (Fig. 1c), the orbital separation will tend to shrink. However, if the orbital separation is too large, evaporation will be weaker and weaker, and generate less and less thrust. If the orbital separation is too small, the companion star could eventually merge with the neutron star.

It is thus critical to be able to modulate the orbital separation to maintain an optimum distance for thrust or deceleration. What could be such a mechanism? It might involve changing the Roche lobe geometry, possibly through accretion or an increase of the magnetic field of the neutron star that would pull or push the companion star. Changing the orbital plane (Fig. 1d) would also disrupt orbital elements, and further stabilization maneuvers may be needed.

Although the stellar density is sparse in the galaxy, performing gravitational assist maneuvers would always be wise to save fuel. The binary stellar engine's ultimate goal is to capture a new star, so it should be able to perform a capture maneuver and eject its depleted companion star when the new star is acquired. In both these maneuvers, one has to take into account the danger of a chaotic dynamics implied by three-body interactions. So the gravitational assist or gravitational capture would have to be wide enough so that the gravitational interaction of the third body does not disturb the close binary.

## 4  SPIDER STELLAR ENGINE: CANDIDATE SYSTEMS

Could our galaxy host a kind of fully steerable binary stellar engine that we proposed? This is a plausible hypothesis in the context of the stellivore hypothesis, that reinterprets some observed accreting binary stars as advanced civilizations feeding on stars (Vidal 2014 [34]; 2016 [16]). The scenario would be the following. For most of its time, a stellivore civilization would eat its home star via accretion. However, energy is never eternal, and instead of eating its star until the end and dying, a stellivore civilization would use its low-mass companion star as fuel not to be *accreted*, but to be *evaporated*, in order to create thrust and travel towards a nearby star. The fundamental *accretion* dynamics of close binaries would be reversed into an *evaporation* dynamics.

This is exactly what seems to happen in a most fascinating subset of close binaries, *spider pulsars*. Spider pulsars are binary stars composed of one millisecond pulsar (MSP) and a very low-mass companion star. If the companion star weights less than 0.1 $M_\odot$, it is called a black widow (BW), if more, typically between 0.1 and 0.7 $M_\odot$, it's a redback (RB, Roberts 2012 [35]). Their phenomenology is exceptional amongst other low-mass X-ray binaries known in the galaxy, in the sense that pulsars in such binary systems usually accrete matter from their companion star, but here they evaporate their companion star instead. We will also discuss the case of transitional MSPs (tMSPs), which alternate between accretion and evaporation phases. The "spider" metaphor comes from female spiders – black widows or redbacks – that eat their male companions after mating. However the metaphor has its limits as the phenomenology is not one of "eating", which would be more appropriate for accretion unto a compact object like in low mass X-ray binaries, but of evaporation. We argue in this section that this peculiar binary configuration might be an instantiation of the binary stellar engine presented in section 3, that we henceforth call the *spider stellar engine* (SSE).

### 4.1  Acceleration

If we assume that there is an evolutionary link between RB and BW pulsars, then we can compute the rocket equation using the difference between the mass of the RB and BW companion. We can use demographics data about spider pulsars (Strader et al. 2019 [36]; Swihart et al. 2022 [37]), an average mass of 1.78 $M_\odot$ for the MSP and an average companion mass of 0.36 $M_\odot$ so we then set the initial mass of the SSE to $m_0$ = 1.78 + 0.36 = 2.16 $M_\odot$. We assume that final mass is the same mass for the MSP, but with an almost evaporated companion of 0.01 $M_\odot$, so we have a final mass of $m_f$ = 1.79 $M_\odot$. The calculation above with equation (1) gave the minimal effective exhaust velocity of $v_{e\,min}$ = 753 km/s. With these numbers, the rocket equation (2) gives a $\Delta v_{min}$ = 141.5 km.s$^{-1}$.

Another way to compute the rocket equation is to look at the evaporation rate. The mass-loss rate has been estimated to be ~10$^{-14}$ $M_\odot$.yr$^{-1}$ for BW PSR J2051–0827 (Stappers et al. 1996 [38]), ~1.9 × 10$^{-14}$ for BW PSR J1544+4937 (Kumari et al. 2023 [39]), ~2 × 10$^{-13}$ for RB PSR J1816+4510 and ~10$^{-12}$ for BW PSR J1959 +2048 (Polzin et al. 2020 [40]), and ~10$^{-10}$ for RB J2215+5135 (Sullivan and Romani 2024 [41])].

A theoretical estimate of ~8 × 10$^{-9}$ for the RB PSR J2039–5617 has been computed, in an attempt to explain its long-term orbital period increase (Clark et al. 2021 [42]). However, the authors rejected the possibility of such a high mass-loss because it would evaporate the companion too quickly (in 19 Myr), and that such a high evaporation rate would be at odds with other estimates.

This relatively low evaporation rate is a central known issue to establish both (1) an evolutionary link between redbacks and black widows, but also (2) the link between spider pulsars and single MSPs, because evaporation by γ-rays is insufficient to transform a typical star into a black widow companion over a Hubble time (Ginzburg and Quataert 2020 [43]; 2021 [44]).

One can also note that the alternance of evaporation and accretion phases may also drive more mass loss, as in the case of tMSPs, where accretion can amount to 1.6 × 10$^{-12}$ to 1.6 × 10$^{-10}$ $M_\odot$.yr$^{-1}$ (Papitto and Martino 2022 [45]).

From the SSE perspective, the lower rates may represent a selection effect because the longest part of a galactic journey is spent cruising, and it is thus rare to observe the more energy intensive acceleration or deceleration phases. We thus predict that we will discover spider pulsars that have radical evaporation rate changes, possibly with a mass loss differential of 2-3 orders of magnitude. As a whole, we also note that the observation driven estimates above span 4 orders of magnitude (from 10$^{-10}$ to 10$^{-14}$ $M_\odot$ yr$^{-1}$), which might reflect a variety of engines and their regimes.

### 4.2  Exhaust velocity

A necessary condition for the SSE hypothesis is that the material is ejected above the escape velocity that we calculated to be 753 km.s$^{-1}$ in our model. Studying PSR J1959+2048, Lin et al. (2023 [46]) did find a wind outflow mean effective velocity of 954 ± 99 km.s$^{-1}$ in the ingress and 604 ± 47 km.s$^{-1}$ in the egress, and thus a weighted average of the two gives 668 ± 42 km.s$^{-1}$. However, another study using the FAST telescope found 2,050.0

km.s$^{-1}$ (Du et al. 2023 [47]). Some other pulsar wind flows have been estimated, such as for PSR J2051−0827 with 540 km.s$^{-1}$ [47], with another estimate one order of magnitude higher of ~5,000 km.s$^{-1}$ by Polzin et al. 2019 [48]. PSR J2055 +3829 was estimated with a wind outflow of ~1,030 km.s$^{-1}$ (Polzin et al. 2019 [48]). The estimates vary and uncertainties remain, but many of these velocities easily reach escape velocities.

We can add two arguments that mass is ejected outside of the gravitational bound of the binary. First, if it ended up being accreted by the neutron star, it would lead at least some neutron stars in spider binaries to become black holes. Indeed redbacks have a MSP of a median mass of 1.78 $M_\odot$, with a median companion star mass of 0.36 $M_\odot$ (Strader et al. 2019 [36]). Added together the whole system is 2.14 $M_\odot$, which is within the range of modern estimates of the maximum Tolman-Oppenheimer-Volkoff limit that would make a neutron star become a black hole (Rezzolla, Most, and Weih 2018 [49]). However, no black hole evaporating a companion star has been observed.

Second, we note that evaporated material may create very-low mass third bodies (see e.g. Burdge et al. 2022 [50]), but such a low mass third body is not enough to account for the 0.1-0.5 $M_\odot$ that are theorized to be lost through the standard evolutionary path going from redback to black widow systems.

To illustrate the capability of the SSE, let us compute the Δv using the rocket equation from the estimates above for PSR J1959+2048. The mass of the pulsar is about 1.8 $M_\odot$ (Romani et al. 2022 [51]), its companion star about 0.03 $M_\odot$ (Arzoumanian, Fruchter, and Taylor 1994 [52]), which totals to $m_0$ = 1.83 $M_\odot$, an outflow velocity of $v_e$ = 2,050.0 km.s$^{-1}$ (Du et al. 2023 [47]), and evaporation rate of 10$^{-12}$ $M_\odot$.yr$^{-1}$ (Polzin et al. 2020 [40]; Du et al. 2023 [47]) that we reduce by an order of magnitude to account for a 10% duty cycle, once per orbit (10$^{-13}$) and a duration of 10.7 Myr, then Δv = 0.001 km.s$^{-1}$, and $a$ = 3.5 × 10$^{-15}$ m.s$^{-2}$. These figures are much lower than our ideal binary stellar engine, but they remain poorly constrained and following the rocket equation, allowing more time, a faster exhaust velocity, a higher duty cycle, or a higher evaporation rate would change these.

### 4.3 Deceleration

From the SSE perspective, black widows have not a lot of propellant left, so they are likely to be in a deceleration phase. Conversely, redbacks have more massive companions, so are likely to be in a cruising or acceleration phase. This is consistent with modeling of the intrabinary shock of spider pulsars (Wadiasingh et al. 2017 [53]; Wadiasingh et al. 2018 [54]; Merwe et al. 2020 [55]), where the shock wraps around the companion in the case of black widows, and around the pulsar in the case of redbacks. These two configurations are consistent with the model of a magnetic sail to improve deceleration (see again Fig. 2a).

To illustrate the plausibility of a deceleration phase, the proper motion of black widow J1641+8049 may have changed in a significant way from the measurement in Lynch et al. 2018 [56], giving μ = 39(3) mas.yr$^{-1}$ to the one in Kirichenko et al. 2023 [57], giving μ = 2.02(10) mas.yr$^{-1}$. Such an important difference demands following up, although at least some of the variance is likely due to the two different telescopes used (see also discussions in Mata Sánchez et al. 2023 [58]; McEwen et al. 2024 [59]).

The proper motion of black widow pulsar PSR J1959+2048 might also have slowed down as measured from μR.A. = -16.0 ± 0.5 mas.yr$^{-1}$ and μdecl. = -25.8 ± 0.6 mas.yr$^{-1}$ (Arzoumanian, Fruchter, and Taylor 1994 [52]) to the recent highly accurate measurement made by Romani et al. [51] of μR.A. = -15.81 ± 0.05 mas.yr$^{-1}$ and μdecl. = -25.54 ± 0.08 mas.yr$^{-1}$. The changes remains within the uncertainties, but repeating a highly accurate Very Large Base Interferometry measurement of its proper motion in a few years should resolve the question of whether its velocity is changing, as we are predicting. The question of when to decelerate ultimately depends on the braking capacity of the SSE. The greater it can brake, the later it can start decelerating.

One can also predict that the BW deceleration will ultimately match the velocity of a nearby target star. Indeed, a capture scenario would require as much as possible matching the inertial frame of the target star, pretty much in the same way as one first needs to run to match a train's speed before attempting to climb on it.

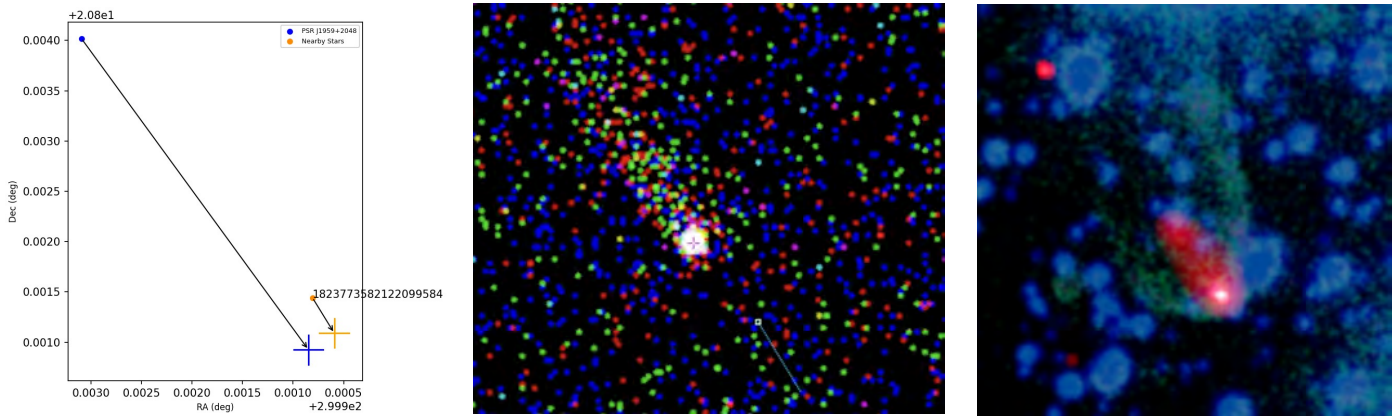

**Fig.3** The original black widow pulsar PSR J1959+2048. Left: the BW pulsar (in blue) is plotted in the RA-DEC plane, and its proper motion vector is displayed until it reaches a close encounter with a target star, in orange, whose Gaia DR3 ID is displayed (see Vidal et al. 2023 [60] for the method used to find close encounters). Middle: a Chandra X-ray view of the BW pulsar, displaying a comet-like tail; the candidate target star is also visible in the bottom right (visualization with ESASky). Right: The composite image on the right shows the X-ray tail (in red/white) and a bow shock visible in the optical (green). *Credit: X-ray: NASA/CXC/ASTRON/B. Stappers et al. [38]; Optical: AAO/J.Bland-Hawthorn & H.Jones; scale: 1.2 arcmin on a side.*

We extrapolated the proper motion of PSR J1959+2048 in the future and searched for close stellar encounters. We used the Gaia catalog to filter only stars that are compatible with the parallax distance of the spider companion (see Fig. 3). We found such a potential target star, Gaia DR3 1823773582122099584 (Vidal et al. 2023 [60]) and that this close encounter may happen as soon as in about 420 years. In the left panel of Fig. 3, we can see that PSR J1959+2048 and the target star have their proper motion vectors aligned, although the pulsar binary has a much higher proper motion. If correct, we predict that the final proper motion of the black widow will match the proper motion of its putative target star, i.e. μR.A. = -1.8504 ± 1.2594 mas.yr$^{-1}$ and μdecl. = 3.0215 ± 1.2457 mas.yr$^{-1}$. Note that the scenario remains uncertain because of the remaining distance uncertainties of both the pulsar and its target star.

In their detailed study of PSR J1959+2048, Romani et. al. (2022 [51]) also noted that the 3D motion of the system appears to be nearly aligned with the spin axis of the MSP. This is not expected to happen (see e.g. the misalignment case of a single MSP in Romani, Slane, and Green 2017 [61]). From the SSE interpretation, this perfect alignment is not only expected, but required to provide maximal thrust since this configuration allows the pulsar wind to blow its maximal power to the companion. As suggested by the authors, alignment can be further checked with scintillation studies (e.g. Ord, Bailes, and van Straten 2002 [62]), and we can add the prediction that all spider pulsars should display such an alignment.

Finally, we would like to note that the target of a SSE might not be systematically towards a nearby star, but in a region where no other MSP is around, possibly for the purpose of adding a new node in the pulsar positioning system (Vidal 2019 [8]). Indeed, spider pulsars are also amongst the fastest spinning MSPs and they are often included in pulsar timing arrays, although redbacks are less good candidates because of their timing noise.

### 4.4 Steering

Regarding steering, the candidate redback 4FGL J1702.7−5655 displays over the years a modulation at a slightly later phase (Corbet et al. 2022 [63]). Such a modulation might be interpreted as a gradual steering in the orbital plane (see Fig. 1b).

We saw that steering to change the orbital plane would require asymmetric heating of the companion as a way to push it outside of the orbital plane (see Fig 1d). Such asymmetric heating has been proposed as a way to better explain peculiar light curves where a direct symmetrical heating model appears insufficient (e.g. Sanchez and Romani 2017 [64]; Stringer et al. 2021 [65]). Although uncertainties regarding inclination can be important, the prediction of a change in inclination can be further tested with future observations and modeling.

The phenomenon of asymmetric heating is often associated with transitional MSPs (tMSPs), which are redbacks that can switch abruptly from an accretion to an evaporation configuration (Papitto and Martino 2022 [45]). The SSE offers three tentative hypotheses:
(1) After the tMSP heats its companion star asymmetrically to perform a steering maneuver outside the orbital plane, it generates an accretion disk to stabilize the new orbital plane. Another stabilization maneuver might be to generate asymmetric heating on the opposite side.
(2) The tMSP is a SSE that has just started its engine, so

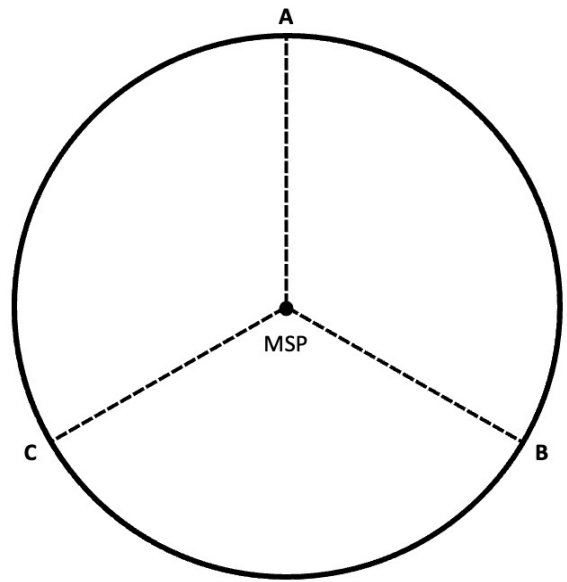

**Fig.4** PSR J1311−3430 has a puzzling 2-hour X-ray periodicity in its 1.5-hour orbital period (Luca et al. 2022 [66]). The MSP is in the center, and the companion's star orbit is illustrated with the solid line. This face-on inclination shows the the X-ray pulses at points A, B, and C (assuming clockwise orbit). Note that this periodicity nicely splits the orbital plane into three equal parts, and this may be an effective stabilizing maneuver.

it blows the companion star particularly hard to get momentum and might switch into the accretion mode to compensate the strong orbital separation increase.
(3) The tMSP charges itself via accretion power as it will have to blow a strong wind for a long time to evaporate the companion.

### 4.5 Other potential features and maneuvers

Another black widow that might have performed a stabilization maneuver is PSR J1311−3430, by generating thrust at three different orbital periods, splitting the orbital plane in three equal parts (see Fig. 4).

Another poorly understood observation is that the average transverse velocity is higher for black widows than it is for redbacks (O'Doherty et al. 2023 [67]). This is not trivial to explain from an astrophysical perspective, as pulsars should get most of their recoil kick velocity from a supernova explosion, and then only continue to decelerate as they age. We have plotted the transverse velocities of spider pulsars, and compared them with single normal pulsars (green) and single MSPs (blue) in Figure 5.

The SSE offers a natural interpretation: black widows have much less propellant to carry, and are thus easier to accelerate and steer. This can be illustrated with the *rocket mass fraction*, i.e. the mass percentage of propellant relative to the whole mass vehicle, see Table 2 (e.g. Pettit 2012 [69]).

Another issue is to explain the long-term orbital period variations that are reported, for example in PSR J1959+2048 (Applegate and Shaham 1994 [70]), J2051-0827 (Shaifullah et al. 2016 [71]) or PSR J1544+4937 (Kumari et al. 2023 [39]). Orbital modulations have no definite explanation (see e.g. Corbet et al. 2022 [63]), although variation in the gravitation-al quadrupole moment of the companion star is often invoked as a mechanism (Applegate 1992 [72]; Applegate and Shaham 1994 [70]).

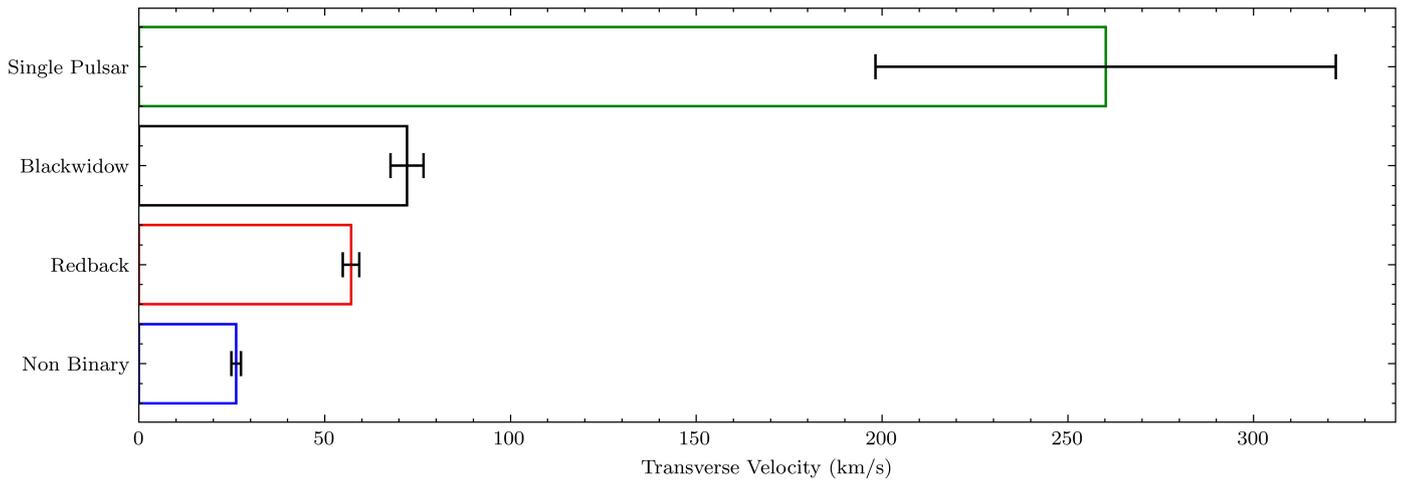

**Fig.5** Average transverse velocity of different classes of pulsars, in km.s$^{-1}$. From top to bottom, we have single normal pulsars (green), black widows (black), redbacks (red) and single MSPs (blue). Note on the data: for spider pulsars, we used demographics of black widows and redbacks (Swihart et al. 2022 [37]), while non-binary (single) MSPs are selected from the galactic field and the single normal pulsars (522) were selected from those with proper motion uncertainties, both from ATNF catalog 1.70 (Manchester et al. 2005 [68]).

The SSE may offer a new interpretation: the strong forces that the MSP exerts on its companion star as it evaporates it would likely tend to extend its orbit. However, to keep maintaining a consistent thrust, the SEE would have to correct its orbital separation and orbital period, otherwise it would generate less and less thrust until no evaporation and thrust is generated. The hypothesis here is that the orbital separation, and more generally the pulsar's magnetic strength, magnetic tilt axis, and orbital elements are tuned to get the desired evaporation rate and other stellar engine requirements. A further implication of the SSE hypothesis is the prediction that drastic phenomenological changes observed in spider binaries originate primarily from the MSP (under control), and not from the companion star (which is just an energy and propellant source).

The recent example of black widow PSR J0610-2100 (Wateren et al. 2022 [73]), which has a low irradiation and does not display strong evidence of long-term orbital variation is consistent with this scenario. In this case, the low irradiation does not push the companion out of orbit, so it doesn't require corrections.

We can also mention the candidate redback PSR J2043+1711 that has been measured to have a significant acceleration of 3.5 ± 0.8 mm.s$^{-1}$.yr$^{-1}$ (Donlon II et al. 2024 [74]). After ruling out many other explanations, the authors conclude that the acceleration is due to a recent close encounter with a nearby star. If the interaction with the third star did really happen in the past, i.e. that is not a foreground or background star, then this is a remarkable event because close encounters in the galactic disk are quite rare (see e.g. Hansen and Zuckerman 2021 [2]).

The author proposes two SSE interpretations. First, that it was a carefully timed maneuver to gain velocity by operating a gravitational assist maneuver during its journey. This may be supported by the fact that the star (Gaia DR3 1811439569904158208) has a low proper motion, and thus that its kinetic energy might have been extracted by the close encounter. The second interpretation is to capture a more massive star in a hierarchical triple, in order for the MSP to change its companion. If this will happen, we might observe an ejection of the current companion, and its replacement by this new star.

## 5   CONCLUSION

We have proposed the first design of a binary stellar engine. Although the binary nature of the stellar engine implies a limited duty cycle in comparison to previous theoretical stellar engine designs, the binary stellar engine has major advantages in terms of maneuverability. Transposing it on a smaller scale, it might also be an inspirational design for advanced propulsion solutions, or for planetary defense purposes such as deflecting asteroids.

Although totally unambiguous evidence of ETI is the ultimate goal in technosignature research, I see the highlighted spider stellar engine candidates and predictions as promising starting points and clues that require further attention, observation, modeling, and follow-up. Spider pulsars thus offer ob-servable stellar engine technosignature candidates, with decades of data, active studies that discover, model and monitor these dazzling systems.

**TABLE 2: Rocket mass fraction for various vehicles, as well as for redbacks and black widows**[*]

| Vehicle | Percent Propellant |
|---|---|
| Car | 4 |
| Locomotive | 7 |
| Fighter jet | 30 |
| Cargo jet | 40 |
| Rocket | 85 |
| Redback / spider stellar engine | 16.8 |
| Black widow / spider stellar engine | 0.55 |

[*] The difference is 30 times less percentage of propellant for black widows compared to redbacks.

## Acknowledgments

The author would very much like to thank the anonymous reviewers who provided major insights that improved the paper, John Hoang and Andy Nilipour for their collaboration to explore the potential goal-directedness of spider pulsars and René Heller for help and discussions about interstellar travel constraints. The Gaia data analysed here are available at the European Space Agency's Gaia archive (https://gea.esac.esa.int/archive).